# Measuring the spatio-temporal field of diffracting ultrashort pulses

P. Bowlan[1]*, M. Lõhmus[2], P. Piksarv[2], H. Valtna-Lukner[2], P. Saari[2] and R. Trebino[1]

[1]*Georgia Institute of Technology, School of Physics 837 State St NW, Atlanta, GA 30332 USA*
[2]*University of Tartu, Institute of Physics, 142 Riia St, Tartu, 51014 Estonia*
* Corresponding author's e-mail address: PamBowlan@gatech.edu

**Abstract**:  Using SEA TADPOLE, we directly measure the spatio-temporal field of diffracting ultrashort pulses with fs-temporal and μm-spatial resolutions.  Using a circular aperture and an opaque disk, we observe boundary wave pulses including their superluminal speeds.

## 1. Introduction

Diffraction phenomena are traditionally studied with monochromatic illumination. But surprisingly it is relatively easy and enlightening to describe diffraction instead using ultrashort pulses, because they can be well localized in both space and time. Also, the elegant, but somewhat forgotten, boundary-wave formulation of diffraction works well for intuitively explaining near-field effects for ultrashort pulses (see references in [1]). Experimentally, it has been difficult to study diffraction of ultrashort pulses in the time domain, however, because it is a spatiotemporal effect. Therefore, to fully characterize a diffracted pulse, it must be measured in both space and time and with fs-temporal and μm-spatial resolutions. Although indirect evidence of the temporal features of diffraction phenomena have been measured [1, 2], direct measurements of the field of diffracted ultrashort pulses have not been reported, to our knowledge.

Here we do so using the ultrashort-pulse measurement technique SEA TADPOLE [3], which records the spatiotemporal electric field $E(x,y,t,z)$ with the appropriate resolutions. We made direct measurements of pulses after propagation through various common optical elements, including annular slits, apertures, and disks. Recently [4], we also used SEA TADPOLE to measure a type of localized wave known as a Bessel-X pulse, which is non-diffracting and has a superluminal group velocity [5]. Our measurements can be viewed as "snapshots in flight," or spatiotemporal slices of the field. Here we show two of our results. First we propagated ultrashort pulses through an opaque disk, making a hole in the beam, and we measured the resulting spatio-temporal field at different distances from the aperture to observe its propagation dependence. This arrangement results in boundary-wave pulses that produce the so-called Spot of Arago. We also propagated the pulses through a circular hole and measured this field, again observing the resulting boundary-wave pulses and their evolution.

## 2. Measurements

A detailed description of SEA TADPOLE can be found in [3]. Briefly, to make our measurements, we sampled a small spatial region of the diffracted field with a single-mode optical fiber (having a mode diameter of 5.4μm, which is therefore our spatial resolution) and then interfered this light with a well-characterized reference pulse in a spectrometer to reconstruct $E(\lambda)$ for that spatial point. Then to find the spatial dependence of the field we scanned the fiber transversely (in *x*) throughout the cross section of the unknown beam, so that $E(\lambda)$ is measured at each *x*, yielding $E(\lambda,x)$. The measured field in the spectral domain can be Fourier transformed to the time domain to give us $E(t,x)$. To measure the *z* (propagation direction) dependence of the spatio-temporal field, the disk or aperture is translated along the propagation direction to bring it nearer or further from the sampling point (the fiber).

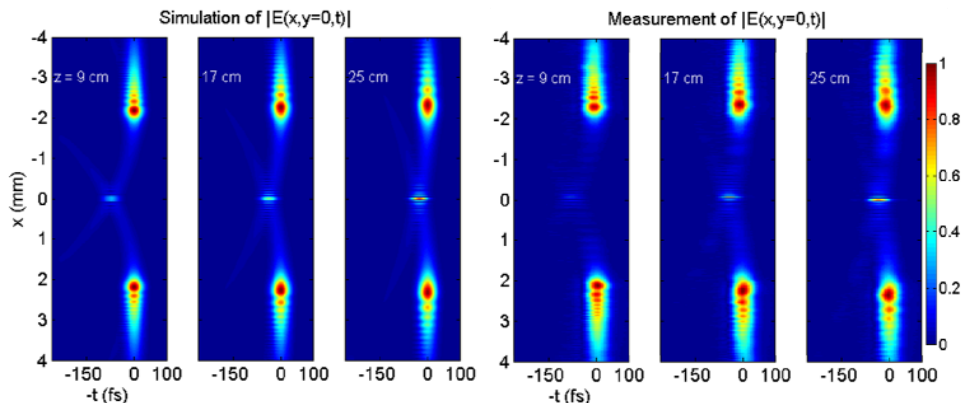

Figure. 1: Measuring the formation and evolution of the spot of Arago behind an opaque disk. Left: Simulations Right: Measurements

We used our KM Labs Ti:Sapphire oscillator, whose pulses had ~37nm of bandwidth (FWHM) and a spot size of 4mm. First we inserted into the beam a 4mm-diameter thin, opaque disk, which made a hole in the beam. We measured $E(\lambda,x)$ at several different distances ($z$'s) after the disk. Three of these measurements are shown in Fig. 1. We performed simulations using the Kirchoff diffraction integral to check our results. These are shown on the left and are in good agreement with our measurements. Because the field behind the disk is very weak, and not much light is coupled into our sampling fiber (because it has a 5.6μm diameter), the measured results are somewhat noisy but the important features are still apparent.

Thinking in terms of the boundary-wave theory of diffraction, the diffracted field can be viewed as the sum of two contributions. One is the undiffracted or main pulse front, which comes from radii greater than that of the disk. This field propagates according to the rules of geometric optics. The other contribution is the boundary wave due to spherical waves emitted from the edge of the disk. These measurements reveal the spatiotemporal structure of the weak boundary waves and the brighter spot at the center of the beam due to their constructive interference, which is known as the Spot of Arago in conventional diffraction theory. Interestingly, our measurements reveal that this spot is delayed in time with respect to the main pulse front and this delay decreases with $z$ indicating a superluminal propagation speed along the $z$ axis (the main pulse front propagates at $c$) which has been observed indirectly in previous studies [2]. This happens, because as $z$ (or the distance from the disk) increases, the extra distance that the boundary waves must propagate (compared to the main pulse front) to reach the $z$ axis ($x = 0$) decreases, so the relative delay of the boundary waves and the bright spot due to their interference decreases. In fact, the group velocity of the Arago spot varies from infinity at $z = 0$ to $c$ for very large values of $z$.

Next, using the same initial field parameters, we propagated the beam through a 4.4mm diameter steel circular aperture and measured the resulting diffraction. The measured fields and simulations at three distances after the aperture are shown in Fig. 2.

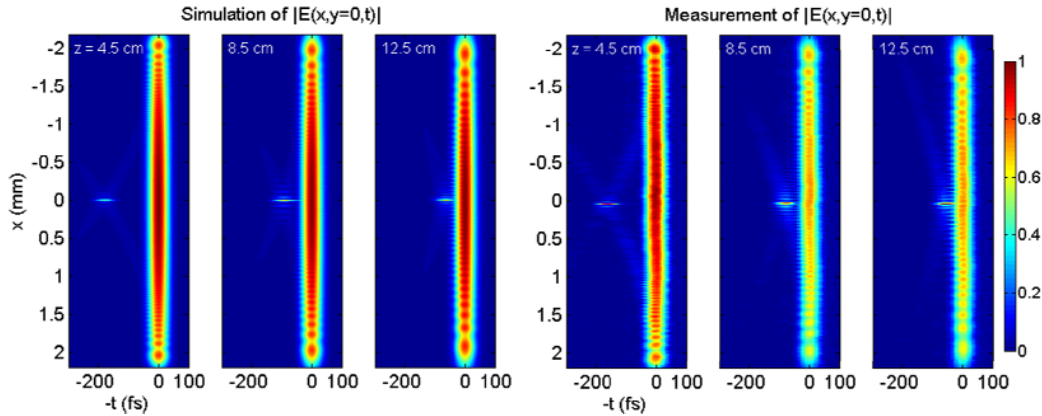

Figure. 2: Formation of the boundary wave pulse after propagation through a circular aperture. Left: Simulations. Right: Measurements.

Again our measurements are in good agreement with the simulations, but with a minor discrepancy in the brightness of the main pulse front which is likely due to the thickness (3.1mm) and imperfect surface quality of our aperture. These measurements show a boundary-wave pulse behind the main pulse-front in time that eventually catches up with it. The boundary-wave pulse in these measurements looks very similar to that shown in Fig. 1. In fact, according to the boundary wave theory of diffraction, because the aperture and disk have similar diameters, their boundary waves are almost the same. So, interestingly, the Arago spot occurs due to any circular boundary and not just a circular disk. The boundary waves in these two measurements look a little different because all of the images are normalized to have maximum of 1, and the main pulse front is much brighter in Fig. 2.

Note that the superluminal speeds of the boundary wave pulses do not contradict relativity, because no information is carried superluminally [5].